\documentclass{JAC2000}
\usepackage{graphicx}
\newcommand{\ud}{\mathrm{d}}
\begin{document}
\title{SPIN-ORBITAL MOTION: SYMMETRY AND DYNAMICS}
\author{A. Fedorova, M. Zeitlin, IPME, RAS, St.~Petersburg, 
V.O. Bolshoj pr., 61, 199178, Russia
\thanks{e-mail: zeitlin@math.ipme.ru}\thanks{http://www.ipme.ru/zeitlin.html;
http://www.ipme.nw.ru/zeitlin.html}}

\maketitle

\begin{abstract}
We present the applications of variational--wavelet 
approach to nonlinear (rational)
model for spin-orbital motion:
orbital dynamics and Thomas-BMT equations for classical spin vector. We
represent the solution of this dynamical system
in framework of periodical wavelets via variational approach 
and multiresolution.

\end{abstract}

\section{INTRODUCTION}

In this paper we consider the applications 
of a new nu\-me\-ri\-cal\--analytical 
technique which is based on the methods of local nonlinear harmonic
analysis or wavelet analysis to the 
spin orbital motion.
Wavelet analysis is a relatively novel set of mathematical
methods, which gives us a possibility to work with well-localized bases in
functional spaces and give for the general type of operators (differential,
integral, pseudodifferential) in such bases the maximum sparse forms.
Our approach in this paper is based on the generalization 
of variational-wavelet 
approach from [1]-[8],
which allows us to consider not only polynomial but rational type of 
nonlinearities [9]. 
The solution has the following form
\begin{equation}\label{eq:z}
z(t)=z_N^{slow}(t)+\sum_{j\geq N}z_j(\omega_jt), \quad \omega_j\sim 2^j
\end{equation}
which corresponds to the full multiresolution expansion in all time 
scales.
Formula (\ref{eq:z}) gives us expansion into a slow part $z_N^{slow}$
and fast oscillating parts for arbitrary N. So, we may move
from coarse scales of resolution to the 
finest one for obtaining more detailed information about our dynamical process.
The first term in the RHS of equation (1) corresponds on the global level
of function space decomposition to  resolution space and the second one
to detail space. In this way we give contribution to our full solution
from each scale of resolution or each time scale.
The same is correct for the contribution to power spectral density
(energy spectrum): we can take into account contributions from each
level/scale of resolution.

In part 2 we consider spin-orbital motion.
In part 3 starting  from  variational formulation we
construct via multiresolution analysis
explicit representation for all dynamical variables in the base of
compactly supported periodized wavelets. 
In part 4 we consider results of numerical calculations.

\section{Spin-Orbital Motion}
Let us consider the system of equations for orbital motion
and Thomas-BMT equation for classical spin vector [10]:
$
\ud q/\ud t={\partial H_{orb}}/{\partial p}, \quad
{\ud p}/{\ud t}=-{\partial H_{orb}}/{\partial q}
$, $\quad\ud s/\ud t=w\times s$,
where
\begin{eqnarray}
H_{orb}&=&c\sqrt{\pi^2+m_0c^2}+e\Phi,\nonumber\\
w=&-&\frac{e}{m_0 c \gamma} (1+\gamma G)\vec B\\
  &+&\frac{e}{m_0^3 c^3\gamma}\frac{G(\vec\pi\cdot\vec B)
\vec\pi}{(1+\gamma)}\nonumber\\
 &+&\frac{e}{m_0^2 c^2\gamma}\frac{G +\gamma G+1}{(1+\gamma)}
[\pi\times E],\nonumber
\end{eqnarray}
$q=(q_1,q_2,q_3), p=(p_1,p_2,p_3)$ are canonical position and momentum,
$s=(s_1,s_2,s_3)$ is the classical spin vector of length $\hbar/2$,
$\pi=(\pi_1,\pi_2,\pi_3)$ is kinetic momentum vector.
We may introduce in 9-dimensional phase space $z=(q,p,s)$ the Poisson brackets
$
\{f(z),g(z)\}=f_qg_p-f_pg_q+[f_s\times g_s]\cdot s
$
and the  Hamiltonian equations are
$
{\ud z}/{\ud t}=\{z,H\}
$
with Hamiltonian
\begin{equation}
H=H_{orb}(q,p,t)+w(q,p,t)\cdot s.
\end{equation}
More explicitly we have
\begin{eqnarray}
\frac{\ud q}{\ud t}&=&\frac{\partial H_{orb}}{\partial p}+\frac{\partial(w\cdot
  s)}{\partial p}\nonumber\\
\frac{\ud p}{\ud t}&=&-\frac{\partial H_{orb}}{\partial q}-\frac{\partial(w\cdot
  s)}{\partial q}\\
\frac{\ud s}{\ud t}&=&[w\times s]\nonumber
\end{eqnarray}
We will consider this dynamical system in [11]
via invariant approach, based on consideration of Lie-Poison structures on
semidirect products.
But from the point of view which we used in [9] we may consider the
similar approximations and then we also arrive to
some type of polynomial/rational dynamics.

\section{Variational Wavelet Approach for Periodic Trajectories}

We start with extension of our approach to the case
of periodic trajectories. The equations of motion corresponding
to our problems may be formulated as a particular case of
the general system of
ordinary differential equations
$
{dx_i}/{dt}=f_i(x_j,t)$, $  (i,j=1,...,n)$, $0\leq t\leq 1$,
where $f_i$ are not more
than rational functions of dynamical variables $x_j$
and  have arbitrary dependence of time but with periodic boundary conditions.
According to our variational approach we have the
solution in the following form
\begin{eqnarray}
x_i(t)=x_i(0)+\sum_k\lambda_i^k\varphi_k(t),\qquad x_i(0)=x_i(1),
\end{eqnarray}
where $\lambda_i^k$ are the roots of reduced algebraical
systems of equations
with the same degree of nonlinearity and $\varphi_k(t)$
corresponds to useful type of wavelet bases (frames).
It should be noted that coefficients of reduced algebraical system
are the solutions of additional linear problem and
also
depend on particular type of wavelet construction and type of bases.

Our constructions are based on multiresolution ap\-pro\-ach. Because affine
group of translation and dilations is inside the approach, this
method resembles the action of a microscope. We have contribution to
final result from each scale of resolution from the whole
infinite scale of spaces. More exactly, the closed subspace
$V_j (j\in {\bf Z})$ corresponds to  level j of resolution, or to scale j.
We consider  a r-regular multiresolution analysis of $L^2 ({\bf R}^n)$
(of course, we may consider any different functional space)
which is a sequence of increasing closed subspaces $V_j$:
\begin{equation}
...V_{-2}\subset V_{-1}\subset V_0\subset V_{1}\subset V_{2}\subset ...
\end{equation}

Then just as $V_j$ is spanned by dilation and translations 
of the scaling function,
so $W_j$ are spanned by translations and dilation of the mother wavelet
$\psi_{jk}(x)$, where
\begin{equation}
\psi_{jk}(x)=2^{j/2}\psi(2^j x-k).
\end{equation}
All expansions, which we used, are based on the following properties:
\begin{eqnarray}
&& L^2({\bf R})=\overline{V_0\displaystyle\bigoplus^\infty_{j=0} W_j}
\end{eqnarray}

We need also to find
in general situation objects
\begin{eqnarray}
\Lambda^{d_1 d_2 ...d_n}_{\ell_1 \ell_2 ...\ell_n}=
 \int\limits_{-\infty}^{\infty}\prod\varphi^{d_i}_{\ell_i}(x)\ud x,
\end{eqnarray}
but now in the case of periodic boundary conditions.
Now we consider the procedure of their
calculations in case of periodic boundary conditions
 in the base of periodic wavelet functions on
the interval [0,1] and corresponding expansion (1) inside our
variational approach. Periodization procedure
gives us
\begin{eqnarray}
\hat\varphi_{j,k}(x)&\equiv&\sum_{\ell\in Z}\varphi_{j,k}(x-\ell)\\
\hat\psi_{j,k}(x)&=&\sum_{\ell\in Z}\psi_{j,k}(x-\ell)\nonumber
\end{eqnarray}
So, $\hat\varphi, \hat\psi$ are periodic functions on the interval
 [0,1]. Because $\varphi_{j,k}=\varphi_{j,k'}$ if $k=k'\mathrm{mod}(2^j)$, we
may consider only $0\leq k\leq 2^j$ and as  consequence our
multiresolution has the form
$\displaystyle\bigcup_{j\geq 0} \hat V_j=L^2[0,1]$ with
$\hat V_j= \mathrm{span} \{\hat\varphi_{j,k}\}^{2j-1}_{k=0}$ [12].
Integration by parts and periodicity gives  useful relations between
objects (9) in particular quadratic case $(d=d_1+d_2)$:
\begin{eqnarray}
\Lambda^{d_1,d_2}_{k_1,k_2}&=&(-1)^{d_1}\Lambda^{0,d_2+d_1}_{k_1,k_2},\\
\Lambda^{0,d}_{k_1,k_2}&=&\Lambda^{0,d}_{0,k_2-k_1}\equiv
\Lambda^d_{k_2-k_1}\nonumber
\end{eqnarray}
So, any 2-tuple can be represented by $\Lambda^d_k$.
Then our second additional linear problem is reduced to the eigenvalue
problem for
$\{\Lambda^d_k\}_{0\leq k\le 2^j}$ by creating a system of $2^j$
homogeneous relations in $\Lambda^d_k$ and inhomogeneous equations.
So, if we have dilation equation in the form
$\varphi(x)=\sqrt{2}\sum_{k\in Z}h_k\varphi(2x-k)$,
then we have the following homogeneous relations
\begin{equation}
\Lambda^d_k=2^d\sum_{m=0}^{N-1}\sum_{\ell=0}^{N-1}h_m h_\ell
\Lambda^d_{\ell+2k-m},
\end{equation}
or in such form
$A\lambda^d=2^d\lambda^d$, where $\lambda^d=\{\Lambda^d_k\}_
{0\leq k\le 2^j}$.
Inhomogeneous equations are:
\begin{equation}
\sum_{\ell}M_\ell^d\Lambda^d_\ell=d!2^{-j/2},
\end{equation}
 where objects
$M_\ell^d(|\ell|\leq N-2)$ can be computed by recursive procedure
\begin{eqnarray}
&&M_\ell^d=2^{-j(2d+1)/2}\tilde{M_\ell^d},\\ 
&&\tilde{M_\ell^k}=
<x^k,\varphi_{0,\ell}>=
\sum_{j=0}^k {k\choose j} n^{k-j}M_0^j,\quad
\tilde{M_0^\ell}=1.\nonumber
\end{eqnarray}
 So, we reduced our last problem to standard
linear algebraical problem. Then we use the methods from [9].
As a result we obtained
for closed trajectories of orbital dynamics  
the explicit time solution (1) in the base of periodized wavelets (10).

\begin{figure}[htb]
\centering
\includegraphics*[width=60mm]{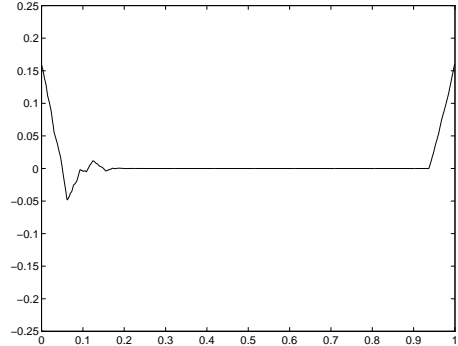}
\caption{Periodic wavelet }
\end{figure}

\section{Numerical Calculations}

In this part we consider numerical illustrations of previous analytical
approach. Our numerical calculations are based on periodic compactly supported
Daubechies wavelets and related wavelet families (Fig.~1).
Also in our modelling we added noise as perturbation to our 
spin orbit configurations.

On Fig.~2 we present according to formulae (2),(6) contributions 
to approximation of our dynamical evolution (top row on the Fig.~3) 
starting from
the coarse approximation, corresponding to scale $2^0$ (bottom row)
to the finest one corresponding to the scales from $2^1$ to  $2^5$
or from slow to fast components (5 frequencies) as details for approximation.
Then on Fig.~3, from bottom to top, we demonstrate the summation
of contributions from corresponding levels of resolution given on
Fig.~2 and as result we restore via 5 scales (frequencies) approximation
our dynamical process(top row on Fig.~3 ).
The same decomposition/approximation we produce also on the level of
power spectral density in the process with noise (Fig.~4).

\begin{figure}[htb]
\centering
\includegraphics*[width=60mm]{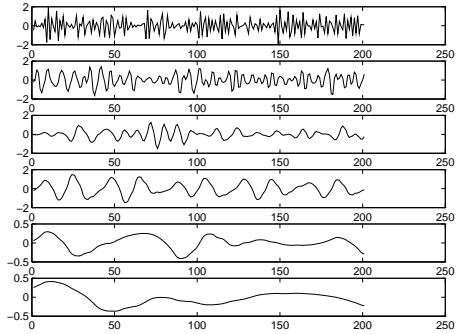}
\caption{Contributions to approximation: from scale $2^1$ to $2^5$ (with noise).}
\end{figure}
\begin{figure}[htb]
\centering
\includegraphics*[width=60mm]{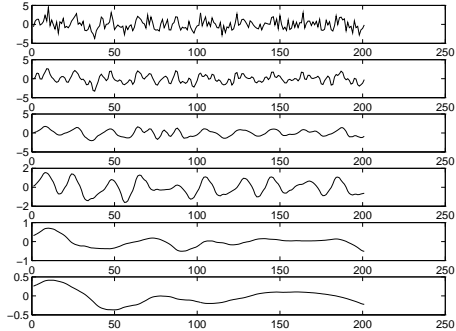}
\caption{Approximations: from scale $2^1$ to $2^5$ (with noise). }
\end{figure}
\begin{figure}[htb]
\centering
\includegraphics*[width=60mm]{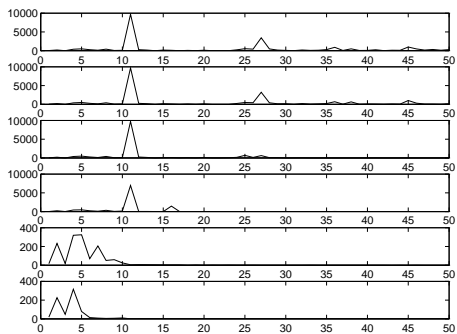}
\caption{Power spectral density: from scale $2^1$ to $2^5$ (with noise)}
\end{figure}

We would like to thank Professor
James B. Rosenzweig and Mrs. Melinda Laraneta for
nice hospitality, help and support during UCLA ICFA Workshop.

 \end{document}